# Analytical solution for free vibrations of simply supported transversally inextensible homogeneous rectangular plate

Milan Batista
University of Ljubljana, Faculty of Maritime Studies and Transport, Slovenia, EU
milan.batista@fpp.uni-lj.si
(January, July 2010)

## Abstract

In article, the exact solution of sinusoidal loaded simply supported elastic transversally inextensible rectangular plate is given. The expressions for displacement and stress components are derived and asymptotic expansion with respect to plate thickness are present. The frequency factors for plate thickness to width ratio 0.01, 0.1, 0.2 and 0.4 and various ratios of plate length to width are given.

The aim of various plate theories is to reduce three dimensional elasticity problems to a two dimensional problems. When theory is developed, one has tempt to compare an exact solution of some simple three dimensional problem to results predicted by the plate theory and in this way obtained direct difference between them. Lack of exact three-dimensional solutions often force authors to compare results only among plate theories or/and among solutions methods and by such comportment, we can hardly decide which plate theory is better.

A typical introducing problem in plate dynamics theories is free vibration of simply supported rectangular plate. The analytical three-dimensional solution of the problem was in 1970 provided by Sirnavis et al (Srinivas, Rao et al. 1970). The three-dimensional numerical solution of the problem was in 1993 given by Liew at al (Liew, Hung et al. 1993). In many articles these solutions are found as benchmark solutions for testing accuracy of various plate theories.

Now, the very basic assumptions in the Mindlin (Mindlin 1951) and various higher-order plate theories (Shufrin and Eisenberger 2005) are that the thickness of plate remains constant during deformation and that a effect of transverse normal stress component is negligible. Physically these assumptions describe a transversally inextensible plate. The numerical accuracy of such plate theories are thus better to be compared with exact analytical three-dimensional solution of transversally inextensible plate. However it seems that such solution is not generally available. The purpose of this article is to fill this gap.

## Governing equations

The object of consideration is a weightless homogeneous transversally isotropic elastic plate bounded by the planes

$$x = y = 0 \qquad x = a \qquad y = b \qquad z = \pm h/2 \tag{1}$$

The stress–strain state of the plate is characterized by the normal stress components $\sigma_x, \sigma_y$, $\sigma_z$, the shear stresses components $\tau_{xy}$, $\tau_{xz}$, $\tau_{yz}$, and the displacement components $u, v, w$. The equations connecting these quantities are equations of motion

$$\frac{\partial \sigma_x}{\partial x} + \frac{\partial \tau_{xy}}{\partial y} + \frac{\partial \tau_{xz}}{\partial z} = \rho \frac{\partial^2 u}{\partial t^2} \qquad \frac{\partial \tau_{xy}}{\partial x} + \frac{\partial \sigma_y}{\partial y} + \frac{\partial \tau_{xy}}{\partial z} = \rho \frac{\partial^2 v}{\partial t^2} \tag{2}$$

$$\frac{\partial \tau_{xz}}{\partial x} + \frac{\partial \tau_{yz}}{\partial y} + \frac{\partial \sigma_z}{\partial z} = \rho \frac{\partial^2 w}{\partial t^2} \tag{3}$$

and the constitutive equations

$$\sigma_x = \frac{2G}{1-\nu}\left(\frac{\partial u}{\partial x} + \nu \frac{\partial v}{\partial y}\right) \qquad \sigma_y = \frac{2G}{1-\nu}\left(\frac{\partial v}{\partial y} + \nu \frac{\partial u}{\partial x}\right) \qquad \tau_{xy} = G\left(\frac{\partial u}{\partial y} + \frac{\partial v}{\partial x}\right) \tag{4}$$

$$\tau_{xz} = G\left(\frac{\partial u}{\partial z} + \frac{\partial w}{\partial x}\right) \qquad \tau_{yz} = G\left(\frac{\partial v}{\partial z} + \frac{\partial w}{\partial y}\right) \tag{5}$$

where $G \equiv \dfrac{E}{2(1+\nu)}$ is the shear modulus, with *E* the modulus of elasticity and $\nu$ Poisson's ratio. The condition of transverse inextensibility is

$$\frac{\partial w}{\partial z} = 0 \tag{6}$$

The boundary conditions of the problem are the following (Srinivas, Rao et al. 1970). Along the sides of the plate the deflection and normal in-plane stress components vanish

$$\begin{aligned} &\text{at} \quad x = 0 \ \text{ and } \ x = a: \quad w = 0 \quad v = 0 \quad \sigma_x = 0 \\ &\text{at} \quad y = 0 \ \text{ and } \ y = b: \quad w = 0 \quad u = 0 \quad \sigma_y = 0 \end{aligned} \tag{7}$$

and on the plate faces the boundary conditions the transverse stress components vanish

$$\text{on} \quad z = \pm h/2: \quad \sigma_z = \tau_{xz} = \tau_{yz} = 0 \tag{8}$$

## Solution

**Form of solution.** The boundary conditions (7) are identically satisfied if the solution is sought in the form of double Fourier series

$$u=\sum_{m=1}^{\infty}\sum_{n=1}^{\infty}U_{mn}(z)\cdot\cos(\alpha_m x)\sin(\beta_n y)\cdot e^{i\omega_{mn}t}\qquad v=\sum_{m=1}^{\infty}\sum_{n=1}^{\infty}V_{mn}(z)\cdot\sin(\alpha_m x)\cos(\beta_n y)\cdot e^{i\omega_{mn}t}$$

$$w=\sum_{m=1}^{\infty}\sum_{n=1}^{\infty}W_{0,mn}\cdot\sin(\alpha_m x)\sin(\beta_n y)\cdot e^{i\omega_{mn}t} \tag{9}$$

where

$$\alpha_m\equiv\frac{m\pi}{a}\qquad\beta_n\equiv\frac{n\pi}{b} \tag{10}$$

and where $U_{mn}(z)$ and $V_{mn}(z)$ are unknown functions, $W_{0,mn}$ are unknown constants and *m* and *n* are integers. By (9) this, the in-plane stress components (4) and (5) become

$$\sigma_x=-\frac{E}{1-\nu^2}\sum_{m=1}^{\infty}\sum_{n=1}^{\infty}\Sigma_{x,mn}\cdot\sin(\alpha_m x)\sin(\beta_n y)\cdot e^{i\omega_{mn}t}\qquad\Sigma_{x,mn}=\alpha_m U_{mn}+\nu\beta_n V_{mn}$$

$$\sigma_y=-\frac{E}{1-\nu^2}\sum_{m=1}^{\infty}\sum_{n=1}^{\infty}\Sigma_{y,mn}\cdot\sin(\alpha_m x)\sin(\beta_n y)\cdot e^{i\omega_{mn}t}\qquad\Sigma_{y,mn}=\beta_n V_{mn}+\nu\alpha_m U_{mn} \tag{11}$$

$$\tau_{xy}=G\sum_{m=1}^{\infty}\sum_{n=1}^{\infty}\mathrm{T}_{xy,mn}\cdot\cos(\alpha_m x)\cos(\beta_n y)\cdot e^{i\omega_{mn}t}\qquad\mathrm{T}_{xy,mn}=\beta_n U_{mn}+\alpha_m V_{mn}$$

$$\tau_{xz}=G\sum_{m=1}^{\infty}\sum_{n=1}^{\infty}\mathrm{T}_{xz,mn}\cdot\cos(\alpha_m x)\sin(\beta_n y)\cdot e^{i\omega_{mn}t}\qquad\mathrm{T}_{xz,mn}=\frac{dU_{mn}}{dz}+\alpha_m W_{0,mn}$$

$$\tau_{yz}=G\sum_{m=1}^{\infty}\sum_{n=1}^{\infty}\mathrm{T}_{yz,mn}\cdot\sin(\alpha_m x)\cos(\beta_n y)\cdot e^{i\omega_{mn}t}\qquad\mathrm{T}_{yz,mn}=\frac{dV_{mn}}{dz}+\beta_n W_{0,mn} \tag{12}$$

Integration of equilibrium equation (3) yield the transverse normal stress component

$$\sigma_z=G\sum_{m=1}^{\infty}\sum_{n=1}^{\infty}\Sigma_{z,mn}\cdot\sin(\alpha_m x)\sin(\beta_n y)\cdot e^{i\omega_{mn}t}\qquad\Sigma_{z,mn}=\alpha_m U_{mn}+\beta_n V_{mn}+\left(\alpha_m^2+\beta_n^2-\frac{\rho\omega_{mn}^2}{G}\right)zW_{0,mn} \tag{13}$$

**Differential equations.** Introducing in-plane stress components (11) into equilibrium equations (2) one obtain a system of ordinary differential equations for unknowns $U_{mn}(z)$ and $V_{mn}(z)$ which may be written in the form

$$\frac{d^2U_{mn}}{dz^2}-\left(\alpha_m^2+\beta_n^2-\frac{1+\nu}{1-\nu}\alpha_m^2-\frac{\rho\omega_{mn}^2}{G}\right)U_{mn}-\frac{1+\nu}{1-\nu}\alpha_m\beta_n V_{mn}=0$$

$$\frac{d^2V_{mn}}{dz^2}-\left(\alpha_m^2+\beta_n^2-\frac{1+\nu}{1-\nu}\alpha_m^2-\frac{\rho\omega_{mn}^2}{G}\right)V_{mn}-\frac{1+\nu}{1-\nu}\alpha_m\beta_n U_{mn}=0 \tag{14}$$

The solution of this system may be written in the form

$$U_{mn}=C_1\beta_n\sinh\left(\tfrac{1}{2}\sqrt{\chi_{mn}^2-\lambda_{mn}^2}\zeta\right)+C_2\alpha_m\sinh\left(\tfrac{1}{2}\sqrt{\frac{2\chi_{mn}^2}{1-\nu}-\lambda_{mn}^2}\zeta\right)$$

$$V_{mn}=-C_1\alpha_m\sinh\left(\tfrac{1}{2}\sqrt{\chi_{mn}^2-\lambda_{mn}^2}\zeta\right)+C_2\beta_n\sinh\left(\tfrac{1}{2}\sqrt{\frac{2\chi_{mn}^2}{1-\nu}-\lambda_{mn}^2}\zeta\right) \tag{15}$$

where

$$\chi_{mn}^2\equiv h^2\left(\alpha_m^2+\beta_n^2\right)\qquad \lambda_{mn}^2\equiv\frac{h^2\rho\omega_{mn}^2}{G} \tag{16}$$

and $\zeta$ is normalized *z* coordinate defined as

$$\zeta\equiv\frac{z}{h/2}\in\left[-1,1\right] \tag{17}$$

The symmetric part of the solution is omitted from the above expressions since symmetrical plate face boundary conditions (8) leads only to zero stress components.

**Characteristic equation.** By means of expressions for transverse stress components (12) and (13) the face boundary conditions (8) yield homogeneous system of three equation for unknown $C_1$, $C_2$ and $W_0$ which may be reduced to the following equations

$$\sqrt{\chi_{mn}^2-\lambda_{mn}^2}\cosh\left(\tfrac{1}{2}\sqrt{\chi_{mn}^2-\lambda_{mn}^2}\right)C_1=0 \tag{18}$$

$$\sqrt{\frac{2\chi_{mn}^2}{1-\nu}-\lambda_{mn}^2}\cosh\left(\tfrac{1}{2}\sqrt{\frac{2\chi_{mn}^2}{1-\nu}-\lambda_{mn}^2}\right)C_3+hW_{0,mn}=0$$

$$2\chi_{mn}^2\sinh\left(\tfrac{1}{2}\sqrt{\frac{2\chi_{mn}^2}{1-\nu}-\lambda_{mn}^2}\right)C_3+\left(\chi_{mn}^2-\lambda_{mn}^2\right)hW_{0,mn}=0 \tag{19}$$

Eq (18) has nontrivial solution if $\alpha^2=\lambda^2$. In this case the system (19) has only trivial solution providing $\nu>-1$ i.e.

$$C_3=0 \qquad W_0=0 \tag{20}$$

When $\alpha^2=\lambda^2$ then (15)

$$U=V=0 \tag{21}$$

If $\alpha^2\neq\lambda^2$ then for nontrivial solution the determinant of the system (19) must be zero. This condition yield the characteristic equation

$$\left(\chi_{mn}^2-\lambda_{mn}^2\right)\sqrt{\frac{2\chi_{mn}^2}{1-\nu}-\lambda_{mn}^2}\cosh\left(\tfrac{1}{2}\sqrt{\frac{2\chi_{mn}^2}{1-\nu}-\lambda_{mn}^2}\right)-2\chi_{mn}^2\sinh\left(\tfrac{1}{2}\sqrt{\frac{2\chi_{mn}^2}{1-\nu}-\lambda_{mn}^2}\right)=0 \tag{22}$$

The solutions of this transcendental equation are characteristic values $\lambda_{mn}$. A values of $\lambda_{mn}$ for various plate dimensions are provided in Table 1-4 and Table 1a-4a where in later the frequency factor is changed to $\lambda\equiv\omega\left(a^2/\pi^2\right)\sqrt{h\rho/D}$. Also on Figure 1 a variation of frequency factor $\lambda_{11}$ with respect to $b/a$ and $h/a$ is shown.

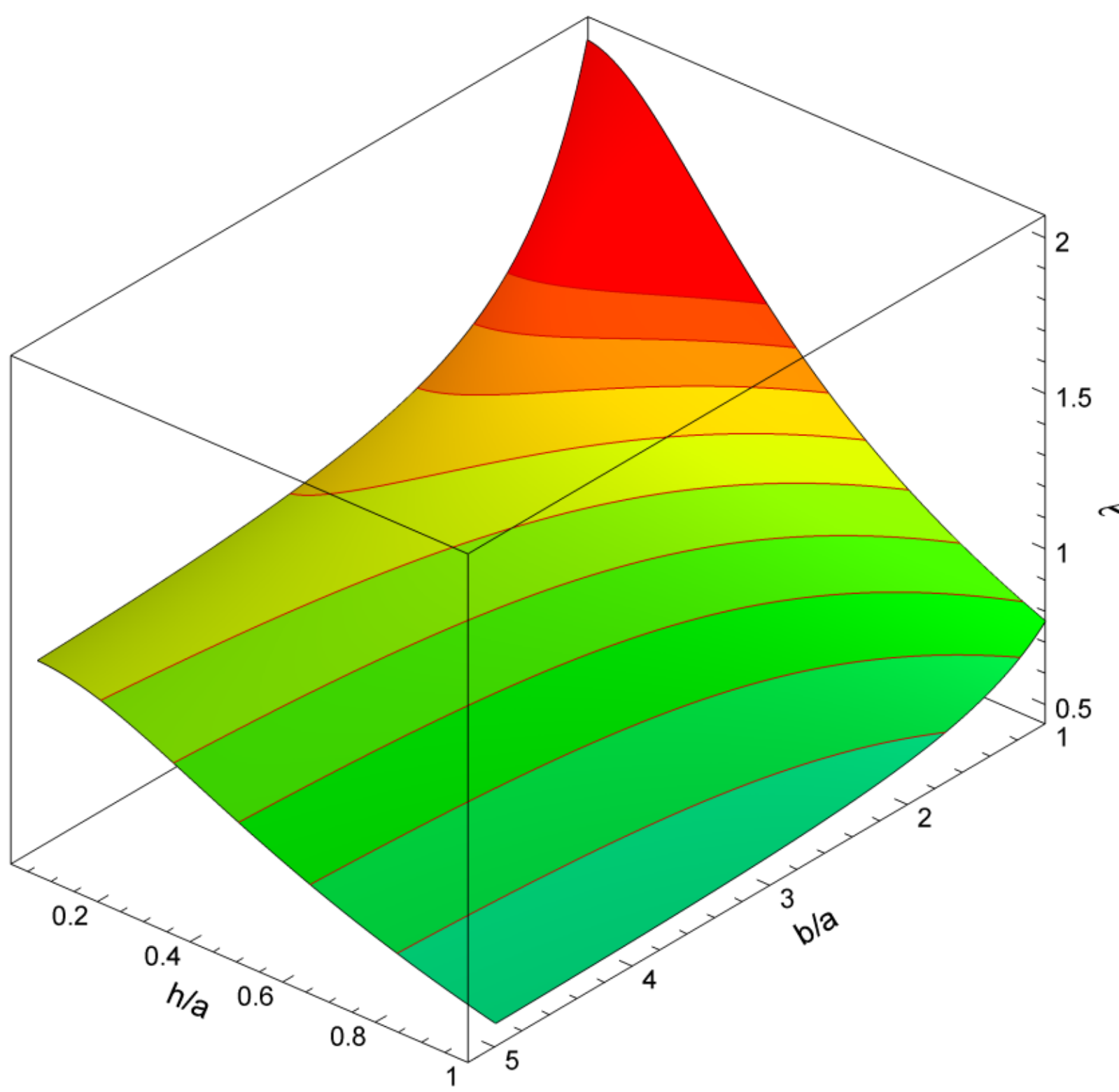

**Figure 1**. Variation of frequency factor $\lambda_{11}$ with respect to $b/a$ and $h/a$

**Table 1.** Frequency factor $\lambda \equiv h\omega\sqrt{\rho/G}$ for $\nu = 0.3$ and $h/a = 0.01$

| | | $b/a$ | | | | | | |
|---|---|---|---|---|---|---|---|---|
| m | n | 1 | 1.5 | 2 | 2.5 | 3 | 5 | 10 |
| 1 | 1 | 0.0010 | 0.0007 | 0.0006 | 0.0006 | 0.0005 | 0.0005 | 0.0005 |
| 1 | 2 | 0.0024 | 0.0013 | 0.0010 | 0.0008 | 0.0007 | 0.0006 | 0.0005 |
| 1 | 3 | 0.0048 | 0.0024 | 0.0016 | 0.0012 | 0.0010 | 0.0007 | 0.0005 |
| 1 | 4 | 0.0082 | 0.0039 | 0.0024 | 0.0017 | 0.0013 | 0.0008 | 0.0006 |
| 2 | 1 | 0.0024 | 0.0021 | 0.0020 | 0.0020 | 0.0020 | 0.0019 | 0.0019 |
| 2 | 2 | 0.0038 | 0.0028 | 0.0024 | 0.0022 | 0.0021 | 0.0020 | 0.0019 |
| 2 | 3 | 0.0062 | 0.0038 | 0.0030 | 0.0026 | 0.0024 | 0.0021 | 0.0020 |
| 2 | 4 | 0.0096 | 0.0053 | 0.0038 | 0.0032 | 0.0028 | 0.0022 | 0.0020 |
| 3 | 1 | 0.0048 | 0.0045 | 0.0044 | 0.0044 | 0.0044 | 0.0043 | 0.0043 |
| 3 | 2 | 0.0062 | 0.0052 | 0.0048 | 0.0046 | 0.0045 | 0.0044 | 0.0043 |
| 3 | 3 | 0.0086 | 0.0062 | 0.0054 | 0.0050 | 0.0048 | 0.0045 | 0.0044 |
| 3 | 4 | 0.0120 | 0.0077 | 0.0062 | 0.0056 | 0.0052 | 0.0046 | 0.0044 |
| 4 | 1 | 0.0082 | 0.0079 | 0.0078 | 0.0078 | 0.0077 | 0.0077 | 0.0077 |
| 4 | 2 | 0.0096 | 0.0085 | 0.0082 | 0.0080 | 0.0079 | 0.0078 | 0.0077 |
| 4 | 3 | 0.0120 | 0.0096 | 0.0088 | 0.0084 | 0.0082 | 0.0079 | 0.0077 |
| 4 | 4 | 0.0153 | 0.0111 | 0.0096 | 0.0089 | 0.0085 | 0.0080 | 0.0078 |

**Table 2.** Frequency factor $\lambda \equiv h\omega\sqrt{\rho/G}$ for $\nu = 0.3$ and $h/a = 0.1$

| | | $b/a$ | | | | | | |
|---|---|---|---|---|---|---|---|---|
| m | n | 1 | 1.5 | 2 | 2.5 | 3 | 5 | 10 |
| 1 | 1 | 0.0930 | 0.0678 | 0.0589 | 0.0547 | 0.0525 | 0.0492 | 0.0478 |
| 1 | 2 | 0.2220 | 0.1276 | 0.0930 | 0.0767 | 0.0678 | 0.0547 | 0.0492 |
| 1 | 3 | 0.4151 | 0.2220 | 0.1482 | 0.1127 | 0.0930 | 0.0639 | 0.0515 |
| 1 | 4 | 0.6525 | 0.3449 | 0.2220 | 0.1615 | 0.1276 | 0.0767 | 0.0547 |
| 2 | 1 | 0.2220 | 0.1989 | 0.1908 | 0.1870 | 0.1849 | 0.1819 | 0.1807 |
| 2 | 2 | 0.3406 | 0.2536 | 0.2220 | 0.2071 | 0.1989 | 0.1870 | 0.1819 |
| 2 | 3 | 0.5208 | 0.3406 | 0.2725 | 0.2399 | 0.2220 | 0.1954 | 0.1840 |
| 2 | 4 | 0.7453 | 0.4550 | 0.3406 | 0.2848 | 0.2536 | 0.2071 | 0.1870 |
| 3 | 1 | 0.4151 | 0.3947 | 0.3876 | 0.3842 | 0.3824 | 0.3798 | 0.3787 |
| 3 | 2 | 0.5208 | 0.4431 | 0.4151 | 0.4019 | 0.3947 | 0.3842 | 0.3798 |
| 3 | 3 | 0.6839 | 0.5208 | 0.4599 | 0.4310 | 0.4151 | 0.3916 | 0.3816 |
| 3 | 4 | 0.8908 | 0.6240 | 0.5208 | 0.4709 | 0.4431 | 0.4019 | 0.3842 |
| 4 | 1 | 0.6525 | 0.6347 | 0.6285 | 0.6256 | 0.6240 | 0.6217 | 0.6208 |
| 4 | 2 | 0.7453 | 0.6770 | 0.6525 | 0.6410 | 0.6347 | 0.6256 | 0.6217 |
| 4 | 3 | 0.8908 | 0.7453 | 0.6917 | 0.6664 | 0.6525 | 0.6320 | 0.6234 |
| 4 | 4 | 1.0783 | 0.8371 | 0.7453 | 0.7013 | 0.6770 | 0.6410 | 0.6256 |

**Table 3.** Frequency factor $\lambda \equiv h\omega\sqrt{\rho/G}$ for $\nu = 0.3$ and $h/a = 0.2$

| m | n | $b/a$ | | | | | | |
|---|---|---|---|---|---|---|---|---|
| | | 1 | 1.5 | 2 | 2.5 | 3 | 5 | 10 |
| 1 | 1 | 0.3406 | 0.2536 | 0.2220 | 0.2071 | 0.1989 | 0.1870 | 0.1819 |
| 1 | 2 | 0.7453 | 0.4550 | 0.3406 | 0.2848 | 0.2536 | 0.2071 | 0.1870 |
| 1 | 3 | 1.2745 | 0.7453 | 0.5208 | 0.4063 | 0.3406 | 0.2399 | 0.1954 |
| 1 | 4 | 1.8580 | 1.0897 | 0.7453 | 0.5627 | 0.4550 | 0.2848 | 0.2071 |
| 2 | 1 | 0.7453 | 0.6770 | 0.6525 | 0.6410 | 0.6347 | 0.6256 | 0.6217 |
| 2 | 2 | 1.0783 | 0.8371 | 0.7453 | 0.7013 | 0.6770 | 0.6410 | 0.6256 |
| 2 | 3 | 1.5411 | 1.0783 | 0.8908 | 0.7978 | 0.7453 | 0.6664 | 0.6320 |
| 2 | 4 | 2.0740 | 1.3767 | 1.0783 | 0.9252 | 0.8371 | 0.7013 | 0.6410 |
| 3 | 1 | 1.2745 | 1.2217 | 1.2029 | 1.1942 | 1.1894 | 1.1825 | 1.1795 |
| 3 | 2 | 1.5411 | 1.3465 | 1.2745 | 1.2404 | 1.2217 | 1.1942 | 1.1825 |
| 3 | 3 | 1.9318 | 1.5411 | 1.3891 | 1.3155 | 1.2745 | 1.2136 | 1.1874 |
| 3 | 4 | 2.4029 | 1.7907 | 1.5411 | 1.4167 | 1.3465 | 1.2404 | 1.1942 |
| 4 | 1 | 1.8580 | 1.8162 | 1.8014 | 1.7945 | 1.7907 | 1.7853 | 1.7830 |
| 4 | 2 | 2.0740 | 1.9156 | 1.8580 | 1.8310 | 1.8162 | 1.7945 | 1.7853 |
| 4 | 3 | 2.4029 | 2.0740 | 1.9500 | 1.8907 | 1.8580 | 1.8097 | 1.7891 |
| 4 | 4 | 2.8142 | 2.2827 | 2.0740 | 1.9724 | 1.9156 | 1.8310 | 1.7945 |

**Table 4.** Frequency factor $\lambda \equiv h\omega\sqrt{\rho/G}$ for $\nu = 0.3$ and $h/a = 0.4$

| m | n | $b/a$ | | | | | | |
|---|---|---|---|---|---|---|---|---|
| | | 1 | 1.5 | 2 | 2.5 | 3 | 5 | 10 |
| 1 | 1 | 1.0783 | 0.8371 | 0.7453 | 0.7013 | 0.6770 | 0.6410 | 0.6256 |
| 1 | 2 | 2.0740 | 1.3767 | 1.0783 | 0.9252 | 0.8371 | 0.7013 | 0.6410 |
| 1 | 3 | 3.2337 | 2.0740 | 1.5411 | 1.2518 | 1.0783 | 0.7978 | 0.6664 |
| 1 | 4 | 4.4429 | 2.8388 | 2.0740 | 1.6434 | 1.3767 | 0.9252 | 0.7013 |
| 2 | 1 | 2.0740 | 1.9156 | 1.8580 | 1.8310 | 1.8162 | 1.7945 | 1.7853 |
| 2 | 2 | 2.8142 | 2.2827 | 2.0740 | 1.9724 | 1.9156 | 1.8310 | 1.7945 |
| 2 | 3 | 3.7914 | 2.8142 | 2.4029 | 2.1938 | 2.0740 | 1.8907 | 1.8097 |
| 2 | 4 | 4.8822 | 3.4488 | 2.8142 | 2.4795 | 2.2827 | 1.9724 | 1.8310 |
| 3 | 1 | 3.2337 | 3.1217 | 3.0817 | 3.0630 | 3.0529 | 3.0381 | 3.0318 |
| 3 | 2 | 3.7914 | 3.3854 | 3.2337 | 3.1614 | 3.1217 | 3.0630 | 3.0381 |
| 3 | 3 | 4.5933 | 3.7914 | 3.4749 | 3.3202 | 3.2337 | 3.1043 | 3.0485 |
| 3 | 4 | 5.5464 | 4.3053 | 3.7914 | 3.5327 | 3.3854 | 3.1614 | 3.0630 |
| 4 | 1 | 4.4429 | 4.3573 | 4.3271 | 4.3130 | 4.3053 | 4.2942 | 4.2895 |
| 4 | 2 | 4.8822 | 4.5603 | 4.4429 | 4.3876 | 4.3573 | 4.3130 | 4.2942 |
| 4 | 3 | 5.5464 | 4.8822 | 4.6303 | 4.5096 | 4.4429 | 4.3442 | 4.3020 |
| 4 | 4 | 6.3727 | 5.3041 | 4.8822 | 4.6758 | 4.5603 | 4.3876 | 4.3130 |

**Table 1a.** Frequency factor $\lambda \equiv \omega\left(a^2/\pi^2\right)\sqrt{h\rho/D}$ for $\nu = 0.3$ and $h/a = 0.01$

| | | $b/a$ | | | | | | |
|---|---|---|---|---|---|---|---|---|
| m | n | 1 | 1.5 | 2 | 2.5 | 3 | 5 | 10 |
| 1 | 1 | 1.9993 | 1.4441 | 1.2497 | 1.1598 | 1.1109 | 1.0398 | 1.0098 |
| 1 | 2 | 4.9955 | 2.7764 | 1.9993 | 1.6395 | 1.4441 | 1.1598 | 1.0398 |
| 1 | 3 | 9.9818 | 4.9955 | 3.2481 | 2.4389 | 1.9993 | 1.3597 | 1.0898 |
| 1 | 4 | 16.9477 | 8.0992 | 4.9955 | 3.5577 | 2.7764 | 1.6395 | 1.1598 |
| 2 | 1 | 4.9955 | 4.4409 | 4.2467 | 4.1569 | 4.1080 | 4.0370 | 4.0071 |
| 2 | 2 | 7.9884 | 5.7717 | 4.9955 | 4.6361 | 4.4409 | 4.1569 | 4.0370 |
| 2 | 3 | 12.9694 | 7.9884 | 6.2429 | 5.4346 | 4.9955 | 4.3565 | 4.0870 |
| 2 | 4 | 19.9276 | 11.0887 | 7.9884 | 6.5522 | 5.7717 | 4.6361 | 4.1569 |
| 3 | 1 | 9.9818 | 9.4283 | 9.2345 | 9.1448 | 9.0960 | 9.0252 | 8.9953 |
| 3 | 2 | 12.9694 | 10.7567 | 9.9818 | 9.6231 | 9.4283 | 9.1448 | 9.0252 |
| 3 | 3 | 17.9414 | 12.9694 | 11.2270 | 10.4202 | 9.9818 | 9.3441 | 9.0750 |
| 3 | 4 | 24.8871 | 16.0641 | 12.9694 | 11.5358 | 10.7567 | 9.6231 | 9.1448 |
| 4 | 1 | 16.9477 | 16.3955 | 16.2022 | 16.1127 | 16.0641 | 15.9934 | 15.9636 |
| 4 | 2 | 19.9276 | 17.7206 | 16.9477 | 16.5899 | 16.3955 | 16.1127 | 15.9934 |
| 4 | 3 | 24.8871 | 19.9276 | 18.1897 | 17.3849 | 16.9477 | 16.3115 | 16.0431 |
| 4 | 4 | 31.8155 | 23.0146 | 19.9276 | 18.4977 | 17.7206 | 16.5899 | 16.1127 |

**Table 2a.** Frequency factor $\lambda \equiv \omega\left(a^2/\pi^2\right)\sqrt{h\rho/D}$ for $\nu = 0.3$ and $h/a = 0.1$

| | | $b/a$ | | | | | | |
|---|---|---|---|---|---|---|---|---|
| m | n | 1 | 1.5 | 2 | 2.5 | 3 | 5 | 10 |
| 1 | 1 | 1.9317 | 1.4082 | 1.2227 | 1.1364 | 1.0894 | 1.0210 | 0.9920 |
| 1 | 2 | 4.6088 | 2.6491 | 1.9317 | 1.5936 | 1.4082 | 1.1364 | 1.0210 |
| 1 | 3 | 8.6188 | 4.6088 | 3.0763 | 2.3397 | 1.9317 | 1.3278 | 1.0691 |
| 1 | 4 | 13.5482 | 7.1609 | 4.6088 | 3.3535 | 2.6491 | 1.5936 | 1.1364 |
| 2 | 1 | 4.6088 | 4.1306 | 3.9614 | 3.8828 | 3.8400 | 3.7777 | 3.7513 |
| 2 | 2 | 7.0731 | 5.2662 | 4.6088 | 4.2997 | 4.1306 | 3.8828 | 3.7777 |
| 2 | 3 | 10.8143 | 7.0731 | 5.6587 | 4.9824 | 4.6088 | 4.0572 | 3.8215 |
| 2 | 4 | 15.4764 | 9.4481 | 7.0731 | 5.9138 | 5.2662 | 4.2997 | 3.8828 |
| 3 | 1 | 8.6188 | 8.1965 | 8.0474 | 7.9782 | 7.9406 | 7.8857 | 7.8626 |
| 3 | 2 | 10.8143 | 9.2014 | 8.6188 | 8.3457 | 8.1965 | 7.9782 | 7.8857 |
| 3 | 3 | 14.2017 | 10.8143 | 9.5504 | 8.9496 | 8.6188 | 8.1318 | 7.9243 |
| 3 | 4 | 18.4963 | 12.9578 | 10.8143 | 9.7776 | 9.2014 | 8.3457 | 7.9782 |
| 4 | 1 | 13.5482 | 13.1803 | 13.0507 | 12.9905 | 12.9578 | 12.9102 | 12.8901 |
| 4 | 2 | 15.4764 | 14.0574 | 13.5482 | 13.3102 | 13.1803 | 12.9905 | 12.9102 |
| 4 | 3 | 18.4963 | 15.4764 | 14.3633 | 13.8371 | 13.5482 | 13.1241 | 12.9437 |
| 4 | 4 | 22.3907 | 17.3815 | 15.4764 | 14.5628 | 14.0574 | 13.3102 | 12.9905 |

**Table 3a.** Frequency factor $\lambda \equiv \omega\left(a^2/\pi^2\right)\sqrt{h\rho/D}$ for $\nu = 0.3$ and $h/a = 0.2$

| | | $b/a$ | | | | | | |
|---|---|---|---|---|---|---|---|---|
| m | n | 1 | 1.5 | 2 | 2.5 | 3 | 5 | 10 |
| 1 | 1 | 1.7683 | 1.3165 | 1.1522 | 1.0749 | 1.0326 | 0.9707 | 0.9444 |
| 1 | 2 | 3.8691 | 2.3620 | 1.7683 | 1.4785 | 1.3165 | 1.0749 | 0.9707 |
| 1 | 3 | 6.6162 | 3.8691 | 2.7036 | 2.1092 | 1.7683 | 1.2456 | 1.0143 |
| 1 | 4 | 9.6453 | 5.6567 | 3.8691 | 2.9209 | 2.3620 | 1.4785 | 1.0749 |
| 2 | 1 | 3.8691 | 3.5144 | 3.3871 | 3.3276 | 3.2951 | 3.2476 | 3.2275 |
| 2 | 2 | 5.5977 | 4.3454 | 3.8691 | 3.6407 | 3.5144 | 3.3276 | 3.2476 |
| 2 | 3 | 7.9999 | 5.5977 | 4.6241 | 4.1413 | 3.8691 | 3.4593 | 3.2810 |
| 2 | 4 | 10.7666 | 7.1466 | 5.5977 | 4.8031 | 4.3454 | 3.6407 | 3.3276 |
| 3 | 1 | 6.6162 | 6.3420 | 6.2446 | 6.1992 | 6.1744 | 6.1384 | 6.1231 |
| 3 | 2 | 7.9999 | 6.9898 | 6.6162 | 6.4392 | 6.3420 | 6.1992 | 6.1384 |
| 3 | 3 | 10.0285 | 7.9999 | 7.2113 | 6.8290 | 6.6162 | 6.2998 | 6.1637 |
| 3 | 4 | 12.4737 | 9.2960 | 7.9999 | 7.3546 | 6.9898 | 6.4392 | 6.1992 |
| 4 | 1 | 9.6453 | 9.4280 | 9.3511 | 9.3154 | 9.2960 | 9.2677 | 9.2557 |
| 4 | 2 | 10.7666 | 9.9442 | 9.6453 | 9.5048 | 9.4280 | 9.3154 | 9.2677 |
| 4 | 3 | 12.4737 | 10.7666 | 10.1227 | 9.8151 | 9.6453 | 9.3947 | 9.2876 |
| 4 | 4 | 14.6092 | 11.8496 | 10.7666 | 10.2388 | 9.9442 | 9.5048 | 9.3154 |

**Table 4a.** Frequency factor $\lambda \equiv \omega\left(a^2/\pi^2\right)\sqrt{h\rho/D}$ for $\nu = 0.3$ and $h/a = 0.4$

| | | $b/a$ | | | | | | |
|---|---|---|---|---|---|---|---|---|
| m | n | 1 | 1.5 | 2 | 2.5 | 3 | 5 | 10 |
| 1 | 1 | 1.3994 | 1.0863 | 0.9673 | 0.9102 | 0.8786 | 0.8319 | 0.8119 |
| 1 | 2 | 2.6917 | 1.7866 | 1.3994 | 1.2008 | 1.0863 | 0.9102 | 0.8319 |
| 1 | 3 | 4.1966 | 2.6917 | 2.0000 | 1.6246 | 1.3994 | 1.0353 | 0.8648 |
| 1 | 4 | 5.7659 | 3.6842 | 2.6917 | 2.1328 | 1.7866 | 1.2008 | 0.9102 |
| 2 | 1 | 2.6917 | 2.4860 | 2.4113 | 2.3762 | 2.3570 | 2.3289 | 2.3169 |
| 2 | 2 | 3.6523 | 2.9624 | 2.6917 | 2.5597 | 2.4860 | 2.3762 | 2.3289 |
| 2 | 3 | 4.9204 | 3.6523 | 3.1184 | 2.8471 | 2.6917 | 2.4538 | 2.3487 |
| 2 | 4 | 6.3360 | 4.4758 | 3.6523 | 3.2178 | 2.9624 | 2.5597 | 2.3762 |
| 3 | 1 | 4.1966 | 4.0513 | 3.9994 | 3.9752 | 3.9620 | 3.9428 | 3.9346 |
| 3 | 2 | 4.9204 | 4.3936 | 4.1966 | 4.1029 | 4.0513 | 3.9752 | 3.9428 |
| 3 | 3 | 5.9612 | 4.9204 | 4.5098 | 4.3089 | 4.1966 | 4.0288 | 3.9563 |
| 3 | 4 | 7.1981 | 5.5874 | 4.9204 | 4.5847 | 4.3936 | 4.1029 | 3.9752 |
| 4 | 1 | 5.7659 | 5.6549 | 5.6156 | 5.5974 | 5.5874 | 5.5730 | 5.5668 |
| 4 | 2 | 6.3360 | 5.9183 | 5.7659 | 5.6942 | 5.6549 | 5.5974 | 5.5730 |
| 4 | 3 | 7.1981 | 6.3360 | 6.0091 | 5.8525 | 5.7659 | 5.6379 | 5.5831 |
| 4 | 4 | 8.2705 | 6.8836 | 6.3360 | 6.0681 | 5.9183 | 5.6942 | 5.5974 |

**Expressions for displacement and stress components.** The nontrivial solution of system of equations (18) and (19) is

$$C_1 = 0 \qquad C_3 = -\frac{hW_{0,mn}}{2\kappa_{mn}\cosh\kappa_{mn}} \tag{23}$$

where

$$\kappa_{mn} \equiv \tfrac{1}{2}\sqrt{\frac{2\chi_{mn}^2}{1-\nu} - \lambda_{mn}^2} \tag{24}$$

With this the expressions for displacement components (15) are

$$U_{mn} = -\frac{W_{0,mn}}{2}h\alpha_m \frac{\sinh(\kappa_{mn}\zeta)}{\kappa_{mn}\cosh\kappa_{mn}} \qquad V_{mn} = -\frac{W_{0,mn}}{2}h\beta_n \frac{\sinh(\kappa_{mn}\zeta)}{\kappa_{mn}\cosh\kappa_{mn}} \tag{25}$$

Using this, the stress components (11) and (13) become

$$\begin{gathered}\Sigma_{x,mn} = \frac{W_{0,mn}}{2h}h^2\left(\alpha_m^2 + \nu\beta_n^2\right)\frac{\sinh(\kappa_{mn}\zeta)}{\kappa_{mn}\cosh\kappa_{mn}} \qquad \Sigma_{y,mn} = \frac{W_{0,mn}}{2h}h^2\left(\beta_n^2 + \nu\alpha_m^2\right)\frac{\sinh(\kappa_{mn}\zeta)}{\kappa_{mn}\cosh\kappa_{mn}} \\ \mathrm{T}_{xy,mn} = -\frac{W_{0,mn}}{h}h^2\alpha_m\beta_n \frac{\sinh(\kappa_{mn}\zeta)}{\kappa_{mn}\cosh\kappa_{mn}}\end{gathered} \tag{26}$$

The transverse stress components are

$$\begin{gathered}\Sigma_{z,mn} = \frac{W_{0,mn}}{2h}\left(\chi_{mn}^2 - \lambda_{mn}^2\right)\left[\zeta - \frac{\sinh(\kappa_{mn}\zeta)}{\sinh\kappa_{mn}}\right] \\ \mathrm{T}_{xz,mn} = \frac{W_{0,mn}}{h}h\alpha_m\left[1 - \frac{\cosh(\kappa_{mn}\zeta)}{\cosh\kappa_{mn}}\right] \qquad \mathrm{T}_{yz,mn} = \frac{W_{0,mn}}{h}h\beta_n\left[1 - \frac{\cosh(\kappa_{mn}\zeta)}{\cosh\kappa_{mn}}\right]\end{gathered} \tag{27}$$

**The expressions for stress resultants.** By definition the stress resultants are

$$M_x \equiv \int_{-h}^{h} \sigma_x z\,dz = -\frac{E}{1-\nu^2}\sum_{m=1}^{\infty}\sum_{n=1}^{\infty} m_{x,mn}\cdot\sin(\alpha_m x)\sin(\beta_n y)\cdot e^{i\omega_{mn}t}$$

$$M_y \equiv \int_{-h}^{h} \sigma_y z\,dz = -\frac{E}{1-\nu^2}\sum_{m=1}^{\infty}\sum_{n=1}^{\infty} m_{y,mn}\cdot\sin(\alpha_m x)\sin(\beta_n y)\cdot e^{i\omega_{mn}t} \quad (28)$$

$$M_{xy} \equiv \int_{-h}^{h} \tau_{xy} z\,dz = G\sum_{m=1}^{\infty}\sum_{n=1}^{\infty} m_{xy,mn}\cdot\sin(\alpha_m x)\sin(\beta_n y)\cdot e^{i\omega_{mn}t}$$

where

$$m_{x,mn} = \frac{h^3 W_{0,mn}}{4\kappa_{mn}^2}\left(\alpha_m^2+\nu\beta_n^2\right)\left(1-\frac{\sinh\kappa_{mn}}{\kappa_{mn}\cosh\kappa_{mn}}\right) \qquad m_{y,mn} = \frac{h^3 W_{0,mn}}{4\kappa_{mn}^2}\left(\beta_n^2+\nu\alpha_m^2\right)\left(1-\frac{\sinh\kappa_{mn}}{\kappa_{mn}\cosh\kappa_{mn}}\right)$$

$$m_{xy,mn} = \frac{h^3 W_{0,mn}}{2\kappa_{mn}^2}\alpha_m\beta_n\left(1-\frac{\sinh\kappa_{mn}}{\kappa_{mn}\cosh\kappa_{mn}}\right) \quad (29)$$

The transverse shear forces are by definition

$$Q_{xz} \equiv \int_{-h}^{h} \tau_{xz}\,dz = G\sum_{m=1}^{\infty}\sum_{n=1}^{\infty} q_{xz,mn}\cdot\cos(\alpha_m x)\sin(\beta_n y)\cdot e^{i\omega_{mn}t}$$

$$Q_{yz} \equiv \int_{-h}^{h} \tau_{yz}\,dz = G\sum_{m=1}^{\infty}\sum_{n=1}^{\infty} q_{yz,mn}\cdot\sin(\alpha_m x)\cos(\beta_n y)\cdot e^{i\omega_{mn}t} \quad (30)$$

where

$$q_{xz,mn} = hW_{0,mn}\alpha_m\left(1-\frac{\sinh\kappa_{mn}}{\kappa_{mn}\cosh\kappa_{mn}}\right) \qquad q_{yz,mn} = hW_{0,mn}\beta_n\left(1-\frac{\sinh\kappa_{mn}}{\kappa_{mn}\cosh\kappa_{mn}}\right) \quad (31)$$

By these the stress components (26)become

$$\Sigma_{x,mn} = \frac{2m_{x,mn}}{h^2}\frac{\kappa_{mn}^2\sinh(\kappa_{mn}\zeta)}{\kappa_{mn}\cosh\kappa_{mn}-\sinh\kappa_{mn}} \qquad \Sigma_{y,mn} = \frac{2m_{y,mn}}{h^2}\frac{\kappa_{mn}^2\sinh(\kappa_{mn}\zeta)}{\kappa_{mn}\cosh\kappa_{mn}-\sinh\kappa_{mn}}$$

$$\mathrm{T}_{xy,mn} = \frac{2m_{xy,mn}}{h^2}\frac{\kappa_{mn}^2\sinh(\kappa_{mn}\zeta)}{\kappa_{mn}\cosh\kappa_{mn}-\sinh\kappa_{mn}} \quad (32)$$

$$\mathrm{T}_{xz,mn} = \frac{q_{x,mn}}{h}\frac{\cosh\kappa_{mn}-\cosh(\kappa_{mn}\zeta)}{\cosh\kappa_{mn}-\sinh\kappa_{mn}/\kappa_{mn}} \qquad \mathrm{T}_{yz,mn} = \frac{q_{y,mn}}{h}\frac{\cosh\kappa_{mn}-\cosh(\kappa_{mn}\zeta)}{\cosh\kappa_{mn}-\sinh\kappa_{mn}/\kappa_{mn}} \quad (33)$$

**Series expansion.** When plate thickness is small i.e $h/a \ll$ one $\kappa_{mn} \ll 1$ obtain the following expansion for displacement components

$$U_{mn} = -\frac{W_{0,mn}}{2} h\alpha_m \left[ \zeta - \frac{\kappa_{mn}^2}{6} \zeta \left( 3 - \zeta^2 \right) + O\left( \kappa_{mn}^4 \right) \right]$$

$$V_{mn} = -\frac{W_{0,mn}}{2} h\beta_n \left[ \zeta - \frac{\kappa_{mn}^2}{6} \zeta \left( 3 - \zeta^2 \right) + O\left( \kappa_{mn}^4 \right) \right] \tag{34}$$

and the following expansion of stress components

$$\Sigma_{x,mn} = \frac{W_{0,mn}}{2h} h^2 \left( \alpha_m^2 + \nu\beta_n^2 \right) \left[ \zeta - \frac{\beta_{mn}^2}{6} \zeta \left( 3 - \zeta^2 \right) + O\left( \beta_{mn}^4 \right) \right]$$

$$\Sigma_{y,mn} = \frac{W_{0,mn}}{2h} h^2 \left( \beta_n^2 + \nu\alpha_m^2 \right) \left[ \zeta - \frac{\beta_{mn}^2}{6} \zeta \left( 3 - \zeta^2 \right) + O\left( \beta_{mn}^4 \right) \right] \tag{35}$$

$$\mathrm{T}_{xy,mn} = -\frac{W_{0,mn}}{h} h^2 \alpha_m \beta_n \left[ \zeta - \frac{\kappa_{mn}^2}{6} \zeta \left( 3 - \zeta^2 \right) + O\left( \kappa_{mn}^4 \right) \right]$$

$$T_{z,mn} = \frac{W_{0,mn}}{12h} \left( \chi_{mn}^2 - \lambda_{mn}^2 \right) \kappa_{mn}^2 \left[ \zeta \left( 1 - \zeta^2 \right) + O\left( \kappa_{mn}^2 \right) \right]$$

$$\mathrm{T}_{xz,mn} = \frac{W_{0,mn}}{2h} h\alpha_m \left[ \kappa_{mn}^2 \left( 1 - \zeta^2 \right) + O\left( \kappa_{mn}^4 \right) \right] \qquad \mathrm{T}_{yz,mn} = \frac{W_{0,mn}}{2h} h\beta_n \left[ \kappa_{mn}^2 \left( 1 - \zeta^2 \right) + O\left( \kappa_{mn}^4 \right) \right] \tag{36}$$

By this the stresses (32) may be approximately expressed as

$$\Sigma_{x,mn} \approx \frac{6m_{x,mn}}{h^2} \zeta \qquad \Sigma_{y,mn} \approx \frac{6m_{y,mn}}{h^2} \zeta \qquad \mathrm{T}_{xy,mn} \approx \frac{6m_{xy,mn}}{h^2} \zeta$$

$$\mathrm{T}_{xz,mn} \approx \frac{3}{2h} q_{xz,mn} \left( 1 - \zeta^2 \right) \qquad \mathrm{T}_{yz,mn} \approx \frac{3}{2h} q_{yz,mn} \left( 1 - \zeta^2 \right) \tag{37}$$

This are well known Reissner's approximation of stress components.

## Concluding Remark

The present results may be used to obtain accuracy of various plate theories that are based on assumption of transversally inextensibility. As the example the Table 5 provide accuracy of frequency factors for first order plate theory (FOPT) and high order plate theory (HOPT) given by Shufrin and Eisenberger (Shufrin and Eisenberger 2005).

**Table 5.** Comparsion of the frequency factor $\lambda \equiv \omega\left(a^2/\pi^2\right)\sqrt{h\rho/D}$ for square plate with $\nu = 0.3$ and $h/a = 0.4$ . Err is relative error in %. Data for FOPT and HOPT are from Table 1 (Shufrin and Eisenberger 2005)

| m | n | Present | FOPT | err % | HOPT | err % |
|---|---|---|---|---|---|---|
| 1 | 1 | 1.3994 | 1.3970 | 0.17 | 1.3996 | 0.01 |
| 1 | 2 | 2.6917 | 2.6771 | 0.54 | 2.6942 | 0.09 |
| 1 | 3 | 4.1966 | 4.1505 | 1.10 | 4.2116 | 0.36 |
| 2 | 2 | 3.6523 | 3.6199 | 0.89 | 3.6609 | 0.24 |
| 2 | 3 | 4.9204 | 4.8521 | 1.39 | 4.9482 | 0.56 |
| 3 | 1 | 4.1966 | 4.1505 | 1.10 | 4.2116 | 0.36 |
| 3 | 2 | 4.9204 | 4.8521 | 1.39 | 4.9482 | 0.56 |
| 3 | 3 | 5.9612 | 5.8537 | 1.80 | 6.0192 | 0.97 |